\documentclass[conference,letterpaper]{IEEEtran}
\ifCLASSINFOpdf
\else
\usepackage[dvips]{graphicx}
\fi
\usepackage[cmex10]{amsmath}
\usepackage{upgreek}
\usepackage{booktabs}
\usepackage{geometry}
\usepackage{epsfig}
\usepackage{latexsym}
\usepackage{multirow}
\usepackage{stfloats}
\usepackage{epstopdf}
\usepackage{color}  
\usepackage{tabularx} 
\usepackage{amssymb}
\usepackage{enumerate}
\graphicspath{{./Figures/}}
\usepackage{color}
\usepackage{bbm}
\usepackage{bm}
\usepackage{cite}
\usepackage{subfigure}
\usepackage{balance}
\usepackage{mathrsfs}
\usepackage{verbatim}
\usepackage{dsfont}
\usepackage{verbatim}
\usepackage{makecell}
\usepackage{algorithm}
\usepackage{algorithmic}
\usepackage{tikz}
\usepackage{diagbox}
\usepackage{caption}
\usepackage[framemethod=tikz]{mdframed}
\usepackage{multicol}
\usepackage{environ}
\usepackage{tikz}
\begin{document}
\title{Frequency-Position-Fluid Antenna Array for Ultra-dense Connectivity in Terahertz Beamforming Systems}
\author{\IEEEauthorblockN{$\textrm{Heyin Shen}^1$, $\textrm{Chong Han}^1$, $\textrm{Hao Liu}^2$, and $\textrm{Tao Yang}^2$}
\IEEEauthorblockA{$^{1}$Shanghai Jiao Tong University, Shanghai 200240, China. E-mail: \{heyin.shen, chong.han\}@sjtu.edu.cn \\$^{2}$Nokia Shanghai Bell, Shanghai 201206, China. E-mail: \{hao.a.liu, tao.a.yang\}@nokia-sbell.com
	}
}  
\markboth{}
\MakeLowercase
\maketitle
\begin{abstract}
\boldmath
The position-fluid antenna (PFA) architecture has become one of the appealing technologies to support ubiquitous connectivity demand in next-generation wireless systems. Specifically, allowing the antenna to adjust its physical position to one of the predefined ports within a fixed region can introduce additional spatial diversity and improve the signal-to-interference-plus-noise ratio (SINR). In addition, frequency diversity is also widely-explored through frequency interleaving in the terahertz (THz) band. However, the operating bandwidth of one antenna is usually limited to 10\% of the central frequency, which imposes a waste of the ultra-broad bandwidth in the THz band. In light of this, a frequency-position-fluid antenna (FPFA) system is proposed in this paper to facilitate ultra-dense connectivity. Specifically, antennas with non-overlapping operating frequency ranges are deployed at the base station (BS) to expand the total available bandwidth and provide frequency domain diversity, while the PFA-enabled users are capable of providing the spatial domain diversity.
The channel model is first derived, based on which a channel correlation-based frequency allocation strategy is proposed. Then, a minimum-projection-based port selection algorithm is developed with singular-value-decomposition (SVD) precoders. Simulation results show that the proposed FPFA architecture exhibits steady performance with an increasing number of users, and outperforms the PFA and the fixed-antenna system in ultra-dense user deployment.
\end{abstract}
\IEEEpeerreviewmaketitle

\section{Introduction}
\label{section_intro}
With multi-GHz continuous bandwidth, terahertz (THz) wireless communication has become an appealing technologies to fulfill the terabit-per-second (Tbps) data rate requirement in future communications~\cite{Challenges,above100G}. To compensate for the huge propagation loss of THz wave, the multiple-input multiple-output (MIMO) antenna architecture is widely investigated to provide a large array gain, and solve the limited communication distance problem~\cite{dynamic,InterIntra}. However, most of the existing MIMO architectures adopt fixed-position antennas with half-wavelength antenna spacing, which fail to fully explore the spatial diversity of the continuous spatial variation of the wireless channel, thus limiting the performance of spatial multiplexing~\cite{modeling}. Furthermore, with increasing number of access devices and the 6G vision to support ubiquitous connectivity, e.g. $10^7 \text{devices}/\text{km}^2$~\cite{opportunistic}, the number of antennas required to provide beamforming gain tend to be extremely large. 

To solve these problems, a position fluid antenna (PFA)-aided system, or movable antenna (MA) system has been proposed \cite{FAS,Movable_elements,movable_antenna1,movable-statistical,Movable_capacity}. Specifically, by utilizing conductive fluid antennas or through flexible cables and control drivers, the PFA is able to change its physical position within a predefined region. Therefore, PFA has the potential to select the position with the best channel condition, thus improving the system performance without introducing a large number of antennas \cite{opportunities}. As PFA and MA have been significantly studied within the field of antenna technology, their recent introduction into communication offers new opportunities for application~\cite{historical}. In~\cite{FAS}, the authors evaluated the outage probability for a single reception PFA in the Rayleigh fading channels, while the authors in~\cite{movable-statistical} jointly optimize the beamforming matrices and the movable antenna selection. To study the performance of PFA in multi-user scenarios, the authors in~\cite{multiple-access} have developed the fluid antenna multiple access scheme to exploit the deep fade of the user interference, thus improving the signal-to-interference-ratio (SINR).
In~\cite{fluid-III}, the channel model of PFA multi-user system has been developed under RIS-aided rich scattering environment, while in~\cite{movable_antenna1}, ZF-based and MMSE-based joint optimization of port selection and beamforming algorithm have been developed to minimize the transmit power. 

In addition to PFA in the spatial domain, the frequency domain diversity can be also improved to support ultra-dense connectivity. With extremely wide bandwidth at THz frequencies, the frequency interleaving (FI) technology can be utilized where the bandwidth is divided into sub-bands and allocated to different user equipment. As a result, the inter-band users can avoid interference by filtering out unwanted carriers. The authors in \cite{cluster-based} proposed a cluster-based multi-carrier bandwidth division multiple access scheme where spatially proximate users are served by the same beam with different sub-carriers. In~\cite{SS-OFDMA}, a spatial-spread orthogonal frequency division multiple access has been developed where beams of different frequencies are directed to users at distinct angles through true-time-delay devices. However, in practice, the antenna bandwidth is usually limited to around 10\% of the central frequency, e.g., $30$~GHz for a central frequency at $0.3$~THz, which significantly hampers the efficient utilization of the THz spectrum. 

In light of this, a frequency-position-fluid antenna (FPFA) array is proposed in this paper to break through the 10\% bandwidth utilization barrier imposed by the antenna bandwidth, while improving diversity in both spatial and frequency domains to support massive device connectivity. In this architecture, the antennas at the BS are divided into $K$ subarrays, each of which operates in a non-overlapping frequency band to serve a group of users. Within each user group, the users can be divided by spatial diversity through the port selection of PFA and hybrid beamforming design. As a result, the entire system can utilize $K$ times the antenna bandwidth, while supporting a user count that is $K$ times the number of RF chains. Based on the FPFA architecture, we first develop a correlation-based user grouping strategy for frequency allocation, and a min-projection-based port selection strategy, based on which the algorithm for beamforming design is proposed to maximize the sum spectral efficiency.

\textit{Notations:} $\mathbf{A}$ is a matrix, $\mathbf{a}$ is a vector and $a$ is a scalar. $\mathbf{A}(a:b,:)$ denotes the matrix composed of the $a^{\mathrm{th}}$ row to the $b^{\mathrm{th}}$ row of $\mathbf{A}$; $\angle{\mathbf{A}}$ denotes the matrix whose elements have the same phase of the corresponding elements in $\mathbf{A}$. $(\cdot)^*$ denote the complex conjugate of a vector; $(\cdot)^{\dagger}$, $(\cdot)^T$ and $(\cdot)^H$ represents the pseudo inverse, transpose and conjugate transpose of a matrix. $| \cdot |$,  $\Vert \cdot \Vert $and $\Vert \cdot \Vert_F$ denotes the modulus, the Euclidean norm and the Frobenius norm. $\operatorname{det} (\cdot)$ represents the determinant. $\mathbb{C}^{m\times n}$ is the set of complex-valued matrices of dimension $m \times n$. 


\begin{figure*}[htbp]
    \centering
    \subfigure[Antenna array architecture at the base station and each user.]{
	\includegraphics[width=3in]{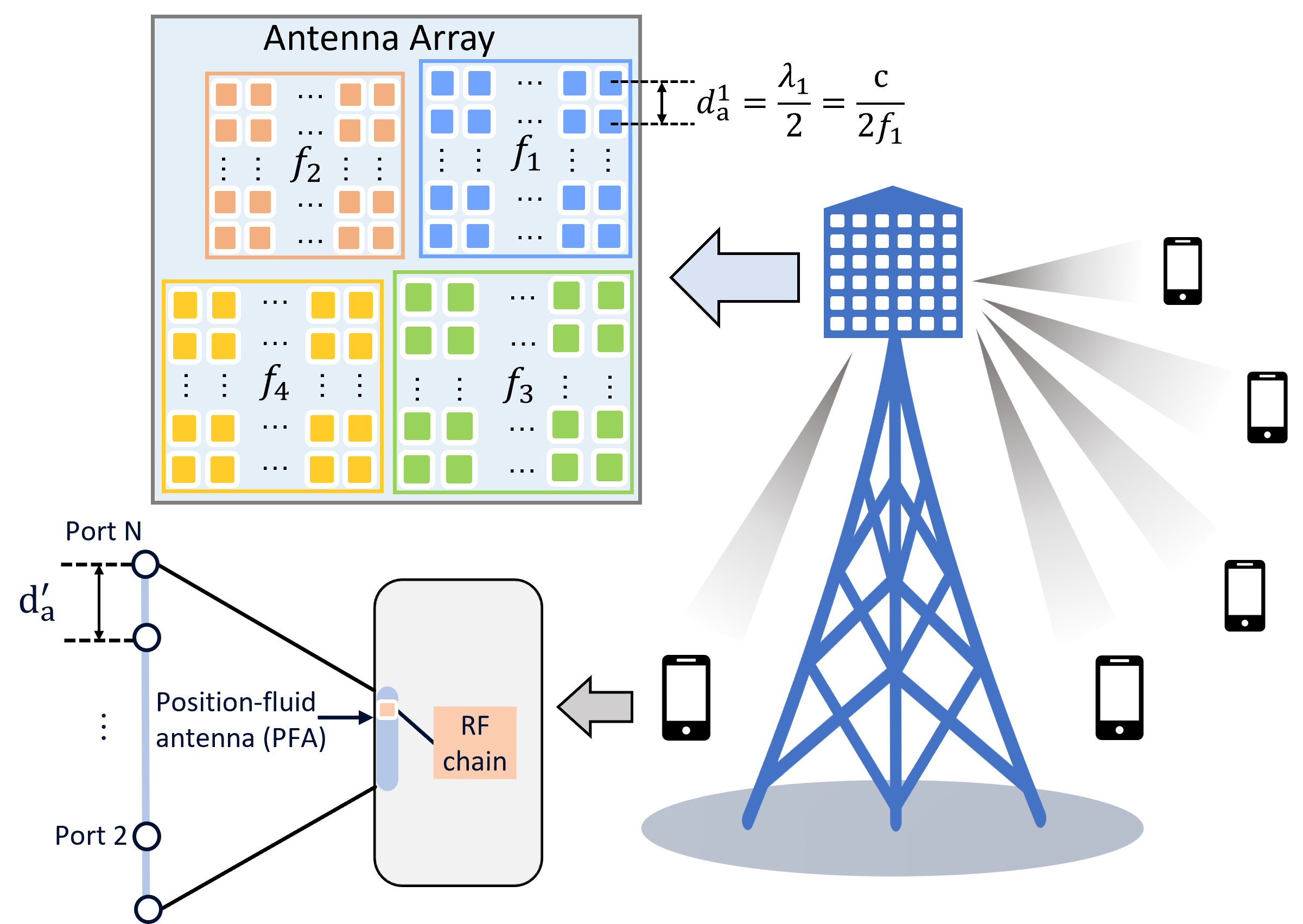}
 \label{fig. antenna array}
    }
    \subfigure[System model for hybrid precoding at the base station.]{
        \includegraphics[width=3in]{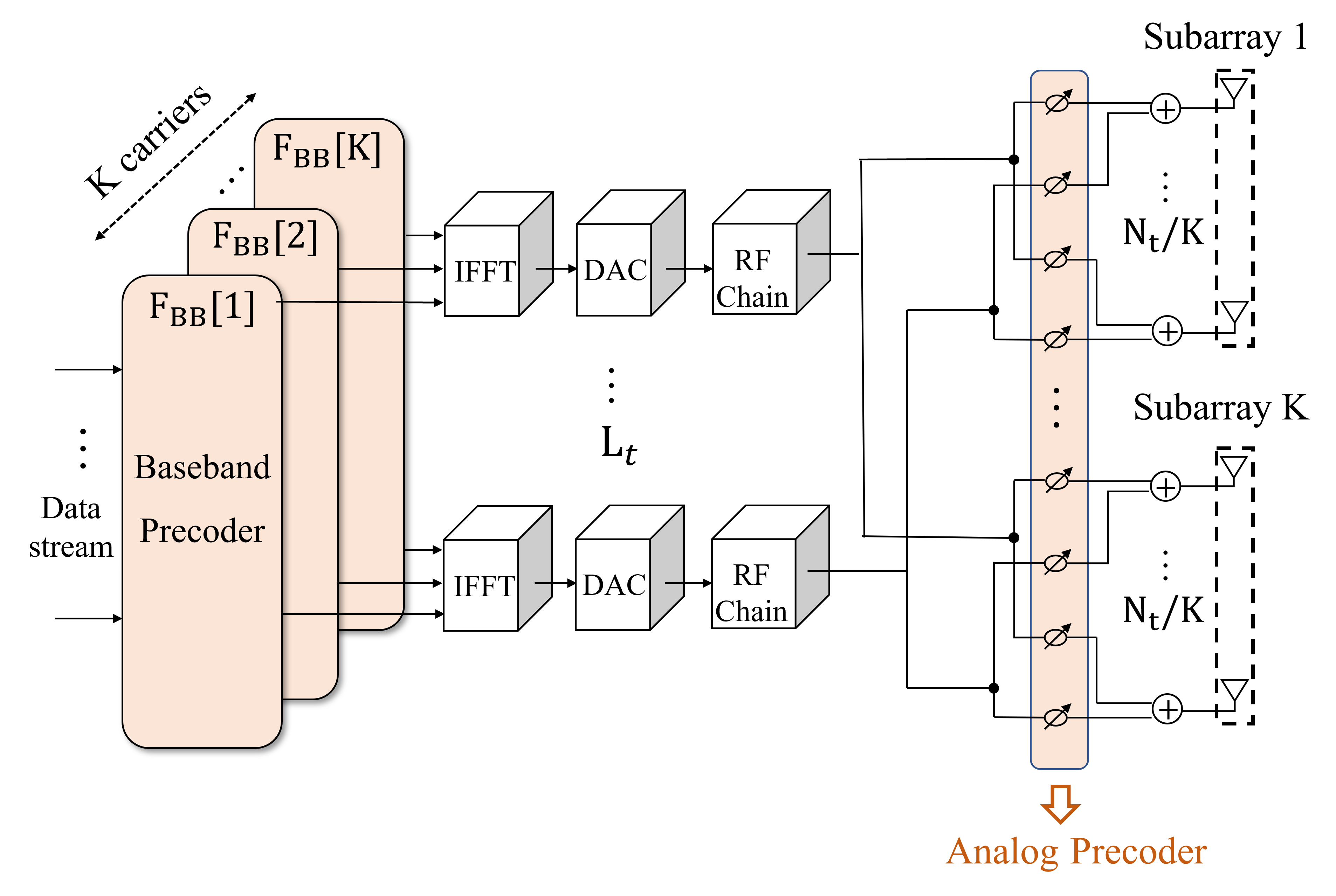}
        \label{fig.precoding}
    }
    \caption{Block diagram of the THz multi-user FPFA hybrid beamforming system.}
    \label{Fig.model}
\end{figure*}

\section{System and Channel Models}
\label{section_Sys_channel_Model}

\subsection{Channel Model}
\label{section_Channel_Model}
Consider a downlink multi-user system where the base station (BS) serves massive users simultaneously. As shown in Fig. \ref{fig. antenna array}, the uniform planar array (UPA) at the BS are divided into $K$ subarrays. Within each subarray, the antennas uses the same central frequency $f_k$ with antenna bandwidth of $0.1f_k$, and the antenna spacing is set as half-wavelength, i.e., $d_a^k = \frac{\lambda_k}{2} = \frac{c}{2f_k}$, where $c$ is the speed of light. While antennas in different subarrays operate in non-overlapping THz bands so that the subarray spacing should be larger than half of the maximum wavelength. At the receiver side, $K$ groups are formed where each group is served by one subarray or one frequency band. The users are equipped with $K$ linear PFAs, each for a different central frequency, but only one of them is activated for the reception of a signal carrier. The PFA can change its position freely to one of the $N$ possible ports that is uniformly distributed along a short line.


By applying the planar-wave-model (PWM), the angle of arrivals (AoAs), angle of departures (AoDs), and the modulus of the complex path gain can be assumed as the same for different ports~\cite{modeling}. While the phase difference of the channel response at each port can be calculated by the difference of the transmission distance as
\begin{equation}
    \mathbf{a}_{rp}^{u}[k] = [1, \cdots, e^{j\frac{2\pi}{\lambda_k}d_p ((N-1)) \sin{\phi^r}}]^T,
\end{equation}
where $\phi^r$ is the receiving angle, and $d_p$ is the the spacing between the ports. Therefore, the pseudo channel matrix between the $k^{\mathrm{th}}$ subarray at the BS and all possible ports at the user can be represented as
\begin{equation}    
\begin{aligned}
    \Tilde{\mathbf{H}}_u[k] = \sum_{p=1}^{N_p}|\alpha_{p}^u[k]|e^{-j \frac{2\pi}{\lambda_k} D_p^u[k]} \mathbf{a}_{rp}^{u}[k](\mathbf{a}_{tp}^{u}[k])^H
\end{aligned}
\end{equation}
where 
\begin{equation}
    \mathbf{a}_{tp}^{u}[k] = [1, \cdots, e^{j\frac{2\pi}{\lambda_k}d_a^k ((X-1))\sin \theta_p \sin{\phi_p}+(Z-1))\cos{\theta_p})}]^T
\end{equation}
refers to the transmit response vector for UPA with $X \times Z$ antenna elements. $|\alpha_p|$ denotes the amplitude of the complex path gain of the $p^{\mathrm{th}}$ path and $D_p^u$ is the distance between the user and the subarray. $\theta$ and $\phi$ refers to the azimuth and elevation angle, respectively. Therefore, for a selected port $n$, the channel matrix is the $n^{\mathrm{th}}$ row of $\Tilde{\mathbf{H}}_u[k]$, i.e., $\mathbf{h}_u = \Tilde{\mathbf{H}}_u[k](n,:)$. Then, the overall channel matrix for all users in group $k$ is represented as $\mathbf{H}[k] = [\mathbf{h}_1^T,\cdots,\mathbf{h}_{U_k}^T]^T$.

\subsection{System Model}
\label{section_system_model}

In FPFA system, the signals are modulated and transmitted across $K$ carriers. Fig. \ref{Fig.model} shows the block diagram of the hybrid precoding architecture where a BS is equipped with $N_t$ antennas to serve $U$ single-antenna users. At the BS, the transmitted symbol at each carrier is $\mathbf{s}[k] = [\mathbf{s}_1^T, \cdots, \mathbf{s}_{U_k}^T]^T \in \mathbb{C}^{U_k \times 1}$ where $U_k$ denotes the number of users served by frequency band $k$. For the $k^{\mathrm{th}}$ carrier, the symbols are processed by the digital precoder $\mathbf{F}_{\mathrm{BB}}[k] \in \mathbb{C}^{L_t \times U_k}$. Then, the signals across the $K$ subcarriers go through a common analog precoder $\mathbf{F}_{\mathrm{RF}} \in \mathbb{C}^{N_t \times L_t}$ shared by all the subcarriers. 
At the receiver, we consider a group of users that are served by subarray $k$. Denote $\mathbf{F}_k \in \mathcal{C}^{N_t/K \times L_t}$ as the sub-analog precoder that connects to subarray $k$, then the overall analog precoder can be written as 
\begin{equation}
    \mathbf{F}_{\mathrm{RF}} = \begin{bmatrix}
        \mathbf{F}_1 \\ \vdots \\ \mathbf{F}_K
    \end{bmatrix}.
\end{equation}

Therefore, the received signal for a typical user in group $k$ can be represented as
\begin{equation}
    \mathbf{y}_u[k] = \mathbf{h}_{u}[k]\mathbf{F}_{k}\mathbf{F}_{\mathrm{BB}}[k]\mathbf{s}[k] + \mathbf{n}_u[k],
\end{equation}
where $\mathbf{n}_u[k]$ denotes the i.i.d complex Gaussian noise for each frequency, i.e., following $\mathcal{CN}(0,\sigma_n^2)$. Thus, the SINR for each user can be written as
\begin{equation}
\label{eq:SINR}
    \gamma_u[k]=\frac{p_u[k]^2\left|\mathbf{h}_u[k] \mathbf{F}_{k} \mathrm{f}_{\mathrm{BB}, u}[k]\right|^2}{\sum_{i\in S_k, i \neq u} p_i[k]^2\left|\mathbf{h}_u[k] \mathbf{F}_{k} \mathbf{f}_{\mathrm{BB}, i}[k]\right|^2+\sigma_n^2},
\end{equation}
where $p_u[k]$ denotes the power allocated to the $u^{\mathrm{th}}$ user at $k^{\mathrm{th}}$ frequency. As a result, the sum spectral efficiency (SE) is represented as
\begin{equation}
    SE = \sum_{k=1}^K\sum_{u \in \mathcal{S}_k}\log_2(1+\gamma_u[k]).
\end{equation}

\subsection{Problem Formulation}
\begin{figure} 
\centering 
\includegraphics[width = 0.35\textwidth]{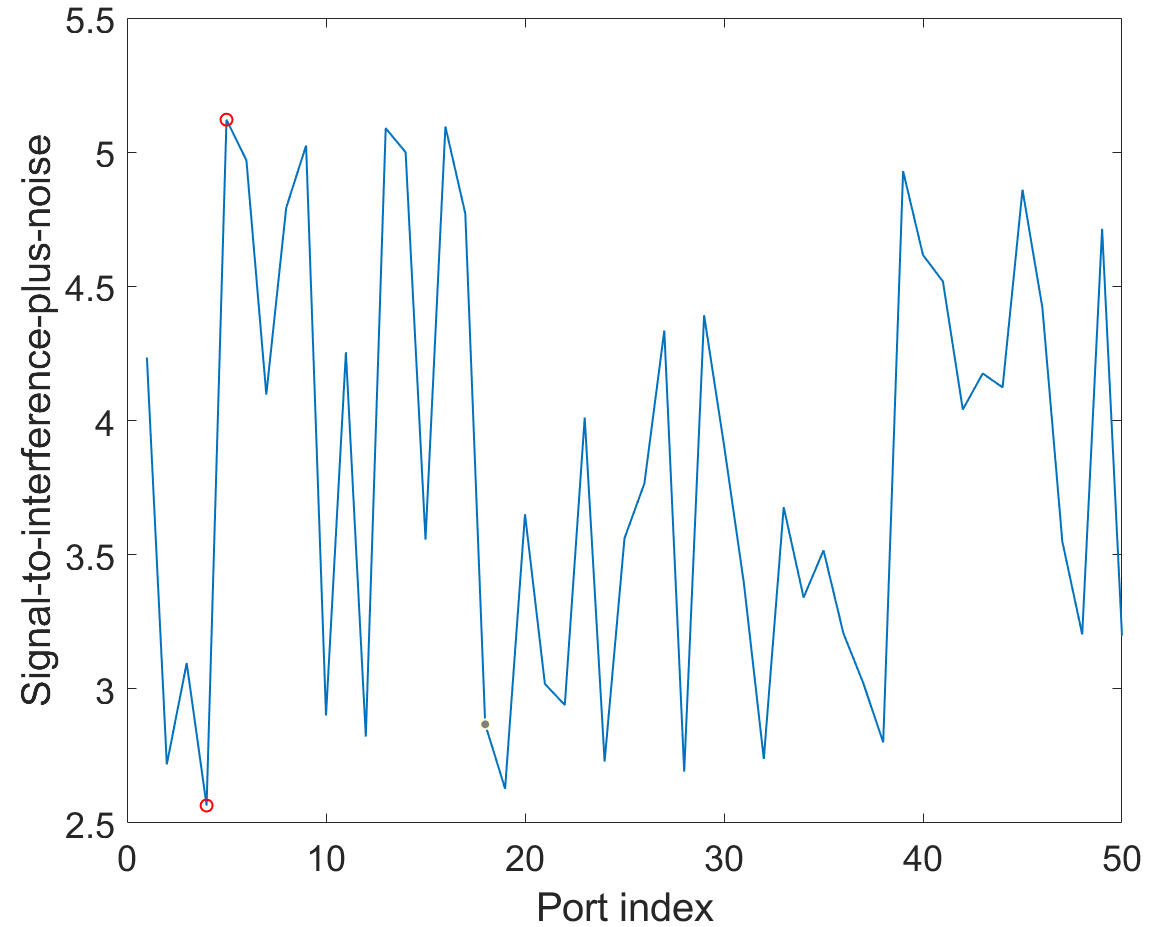} 
\caption{SINR versus the port index.} 
\label{Fig. interference} 
\end{figure}
In the FPFA system, the port selection has a great impact on the SINR of each user. As shown in Fig. \ref{Fig. interference}, we plot the SINR variation versus the port index where we consider 64-antenna BS serving two users for illustration, each of which is equipped with one PFA that can vary its position within a range of $15\lambda$. The results show that the SINR at the best port is over twice that at the worst port. That is to say, by altering the port, we can improve the transmit channel state information. Therefore, it is motivated to consider the joint optimization of the port selection problem and the frequency allocation problem to obtain the best transmit channel matrix for each user, based on which the beamforming matrices are designed to maximize the sum SE. Thus, the overall problem can be written as
\begin{subequations}
    \begin{align}
    \max_{\{\mathbf{h}_u[k]\},\{\mathbf{F}_{\mathrm{RF}}\},\atop \{\mathbf{F}_{\mathrm{BB}}[k]\},\{p_u\}} & SE \\
\text { s. t. } &\sum_{u=1}^{U} p_u^2 \leq P_t, \\
&\left|\mathbf{F}_{k}(i,j)\right|=\frac{1}{\sqrt{N_t/K}}, \forall i, j,k, \label{eq:constant-modulus}\\
& \left\|\mathbf{F}_{\mathrm{RF}} \mathbf{F}_{\mathrm{BB}}[k]\right\|_F^2=U, \forall k, \label{eq:normalized-power}
    \end{align}
\end{subequations}
where Eq. (\ref{eq:constant-modulus}) represents the constant modulus constraint since the analog precoder is implemented by phase shifters, and Eq. (\ref{eq:normalized-power}) represents the normalized power constraint. The optimization of the variable $\mathbf{h}_u[k]$ in fact represents the port selection and the frequency allocation problem.

However, this is a non-convex problem requiring joint optimization over the $K$ carriers due to the frequency allocation and the constant-modulus constraint of the analog beamformer $\mathbf{F}_{RF}$. To make the problem more tractable, we decompose the problem into three sub-problems, through which we first deal with the frequency allocation, and then the optimization of the port selection and beamformers.

\section{Design of Frequency Allocation and Port Selection}
\subsection{Frequency Allocation}
\label{frequency-allocation}
We first consider the frequency allocation problem. Since users allocated with different frequencies do not interfere with each other in FPFA, we aim at minimizing the sum user interference within a user group. However, directly computing the interference requires the knowledge of the beamforming matrix. Therefore, to decouple the frequency allocation subproblem, we use the sum channel correlation coefficient as a measure of the severity of the interference \cite{adaptive_grouping,performance}.

Specifically, the correlation coefficient of two users that are allocated with the same carrier frequency can be defined as 
\begin{equation}
    \rho_{ij} = \frac{\mathbf{h}_i\mathbf{h}^H_j}{\Vert\mathbf{h}_i\Vert \Vert\mathbf{h}_j\Vert}.
\end{equation}
Then, the optimization problem of the frequency allocation can be formulated as
\begin{equation}
\begin{aligned}
    \min_{\{S_k\}} & \sum_{k=1}^K \sum_{i,j \in S_k \atop i \neq j} \rho_{ij}, \\
    \text{subject to} & \cup \mathcal{S}_k = \{1,2,\cdots,U\}, \\
    & S_i \cap S_j = \emptyset, \forall i \neq j.
\end{aligned}
\end{equation}

We now state the channel correlation coefficient in the FPFA system only relates to the angle of the users such that the user grouping scheme can be optimized regardless of the frequency

\textit{Proof:}
We prove it by considering the channel with only one line-of-sight (LoS) path. Cases involving more multipath components are similar.

Given the port selected at the users, the channel matrix for the $m^\mathrm{th}$ and $n^\mathrm{th}$ users with elevation angles and azimuth angles $\theta_m,\theta_n$ and $\phi_m, \phi_m$ can be written as
\begin{equation}
\begin{aligned}
    \mathbf{h}_m & = |\alpha^m|e^{-j\frac{2\pi}{\lambda_m}D^m}\mathbf{a}_{t}^{m},\\
    \mathbf{h}_n & = |\alpha^n|e^{-j\frac{2\pi}{\lambda_n}D^n}\mathbf{a}_{t}^{n}.\\
\end{aligned}
\end{equation}
Therefore, the correlation coefficient is
\begin{equation}
\begin{aligned}
\label{eq:correlation_coefficient}
     \rho_{ij} & = \frac{\lvert\alpha^m \alpha^n e^{-j\frac{2\pi}{\lambda_m}D^m}e^{j\frac{2\pi}{\lambda_n}D^n}\mathbf{a}_{t}^{m}(\mathbf{a}_{t}^{n})^*\rvert}{N_t|\alpha^m||\alpha^n|}\\
     & = \frac{1}{N_t}\lvert \mathbf{a}_{t}^{m}(\mathbf{a}_{t}^{n})^*\rvert \\
     & = \frac{1}{N_t}\lvert 1 + ... +  e^{j\frac{2\pi}{\lambda_m}d_a^m((x-1)\sin\theta_m\cos\phi_m + (z-1)\cos\theta_m}\\
     &\ \ \times e^{j\frac{2\pi}{\lambda_n}d_a^n((x-1)\sin\theta_n\cos\phi_n + (z-1)\cos\theta_n} + ...\rvert
\end{aligned}
\end{equation}
Since the FPFA architecture fixes the antenna spacing for each subarray as $d_k = \frac{\lambda_k}{2}, \forall k$, then \eqref{eq:correlation_coefficient} boils down to
\begin{equation}
\begin{aligned}
    \rho_{ij} & = \frac{1}{N_t}\lvert 1 + \cdots + \\
    & e^{j\pi((x-1)(\sin\theta_m\cos\phi_m+\sin\theta_n\cos\phi_n) + (z-1)(\cos\theta_m+\cos\theta_n))}\rvert,
\end{aligned}
\end{equation}
which only relates to the angle of the users. Thus, we conclude that the user grouping scheme can be optimized regardless of the allocated frequency. \hfill $\blacksquare$

Based on this conclusion, we propose a greedy-based algorithm to form $K$ user groups. We first assume all users are in the same group. Then we greedily find the user pairs with the largest $\rho_{ij}$ and remove one of them from the current group. Next, we search the $K$ groups to find a new group for this user based on the criterion that the resulting sum interference is the smallest. Therefore, the frequency allocation algorithm can be summarized as follows:
\begin{itemize}
\item Step 1: Calculate the correlation coefficient for each user pair $\{\rho_{ij}\}$ and sort them in the descending order as $\{ \rho_1, \rho_2,...,\rho_m,...\rho_{U(U-1)/2}\}$. Set $m = 1$.
\item Step 2: Find the user pairs with coefficient $\rho_m$. If they are in the same group, pull one of the users $u$ out of the current group. 
\item Step 3: Select a new group such that the sum correlation coefficient is the smallest after adding $u$ into the group. Set $m = m+1$, and go back to Step 2 until $m = U(U-1)/2+1$.
\end{itemize}

\subsection{Port Selection}
After the users are allocated with a certain carrier frequency, we next consider the optimization of the port selection and the hybrid precoding scheme. Typically, alternating optimization of the matrices tends to result in a local optimum due to the mismatch between the channel and the precoder \cite{movable_antenna1} and may lead to extremely high computational complexity. Instead, we design the port selection algorithm separately from the beamforming procedure.

Since the channel vectors can now be changed, we aim to suppress the inter-user interference within each group by selecting channel matrices that are orthogonal to each other, thus maximizing the sum SE. However, the choice of the channel vectors at each user is limited by the number of ports so that the optimal channel vector may not be available. Therefore, we propose to select the channel vector that has the minimum projection on other users' channels.

In line with this idea, we denote the projection space for user $u$ as $\bar{\mathbf{H}}_u = [\mathbf{h}_1^T,...,\mathbf{h}_{u-1}^T,\mathbf{h}_{u+1}^T,\cdots,\mathbf{h}_U^T]$. Then, the projection of the channel vector at the $i^{\mathrm{th}}$ port on other users' channel space can be represented as
\begin{equation}
    \mathbf{h}_{proj,u}^i = \bar{\mathbf{H}}_u(\bar{\mathbf{H}}_u^H\bar{\mathbf{H}}_u)^\dagger\bar{\mathbf{H}}_u^H(\mathbf{h}_u^i)^T
\end{equation}
Then, the port index is selected by computing $i^\mathrm{opt} = \min_i \lVert \mathbf{h}_{proj,u}^i \rVert$. To reduce the computational complexity, we conduct this procedure for all users sequentially without iterations. Therefore, the port selection algorithm is summarized as follows.
\begin{itemize}
\item Step 1: Select the first port for the first user and construct the current channel matrix $\bar{\mathbf{H}}$. Set $m = 2$.
\item Step 2: For the $m^{\mathrm{th}}$ user, find the channel vector $\mathbf{h}_m$ with the minimum $\lVert \mathbf{h}_{proj,m}^i \rVert$. 
\item Step 3: Reconstruct the projection space as $\bar{\mathbf{H}} = [\bar{\mathbf{H}},\mathbf{h}_m^T]$. Set $m = m+1$, and go back to Step 2 until $m=U_k$.
\end{itemize}

\subsection{Hybrid Precoding}
After finishing the frequency allocation and port selection, we turn to optimize the hybrid beamformers. For the digital precoder, we apply the block diagonalization (BD) method to cancel the residue interference within a user group~\cite{HBD,HBF}. Specifically, the digital precoder for the $u^\mathrm{th}$ user lies in the null space of other users' channel matrix, given by
\begin{equation}    \mathbf{h}_i[k]\mathbf{F}_{k}\mathbf{f}_{\mathrm{BB,u}}[k] \approx 0, \forall i \neq u.
\end{equation}
Next, the optimal analog precoder for group $k$ can be calculated through SVD such that $\mathbf{F}_{k}^{\mathrm{opt}} = \mathbf{V}(:,1:L_t)$ where $\mathbf{V}$ is the right singular matrix of $\mathbf{H}_k$. To satisfy the constant modulus constraint, the analog precoder can be found by minimizing the Euclidean distance, i.e.,
\begin{equation}
    \min_{\mathbf{F}_{k}} \lVert  \mathbf{F}_{k}^{opt}-\mathbf{F}_{k}\rVert_F^2.
\end{equation}
Therefore, the analog precoder can be expressed as $\mathbf{F}_{k} = \frac{1}{\sqrt{N_t/K}} e ^{j \angle{\mathbf{F}_{k}^{opt}}}$.

\section{Performance Evaluation}
\label{sectin_performance}

\begin{figure*}
\centering
\begin{minipage}[t]{0.48\textwidth}
\centering
\includegraphics[height=0.6\textwidth]{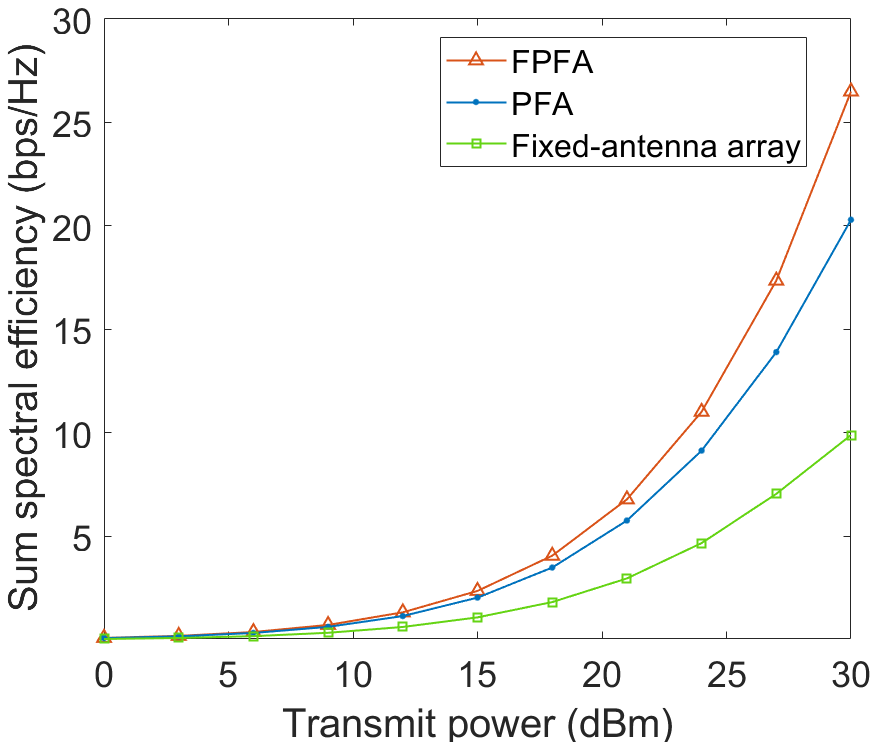}
\caption{Sum SE versus transmit power.}
\label{Fig. power}
\end{minipage}
\begin{minipage}[t]{0.48\textwidth}
\centering
\includegraphics[height=0.6\textwidth]{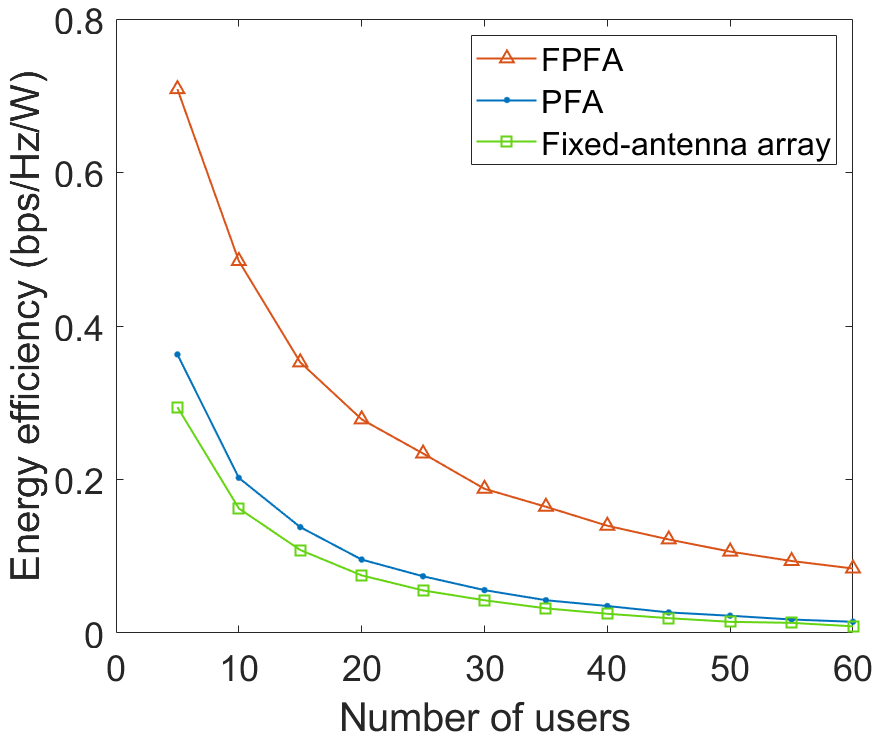}
\caption{Energy efficiency versus the number of users.}
\label{Fig. EE}
\end{minipage}
\end{figure*}

\begin{figure*}
\centering
\begin{minipage}[t]{0.48\textwidth}
\centering
\includegraphics[height=0.6\textwidth]{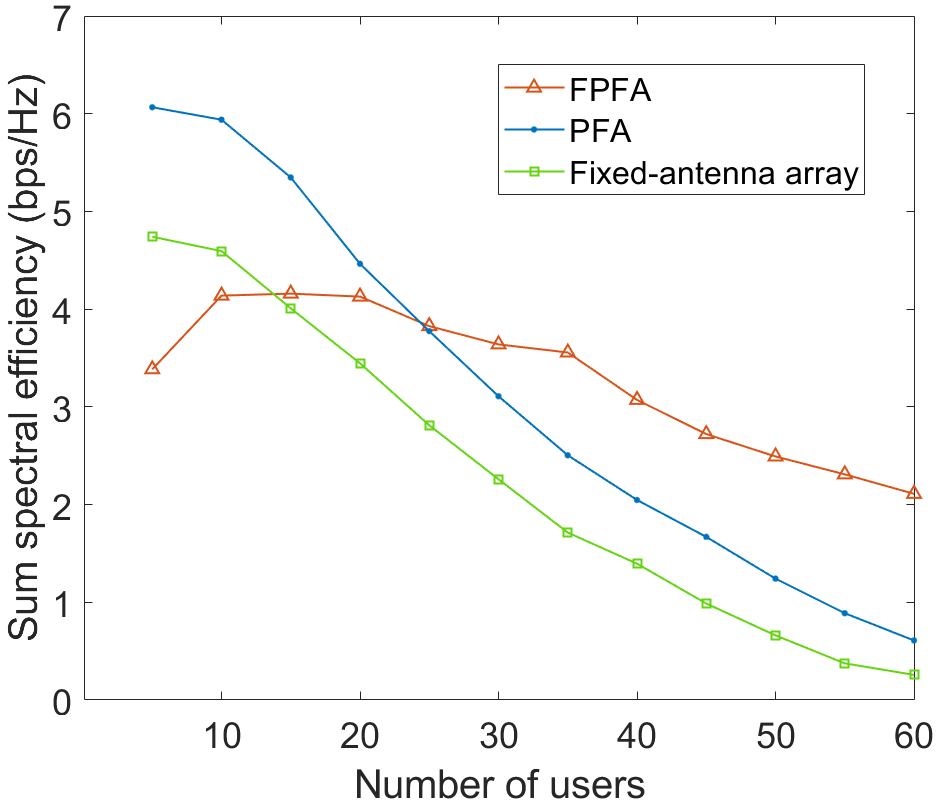}
\caption{Sum SE versus the number of users: BS with 128 antennas.}
\label{Fig. 128 antennas}
\end{minipage}
\begin{minipage}[t]{0.48\textwidth}
\centering
\includegraphics[height=0.6\textwidth]{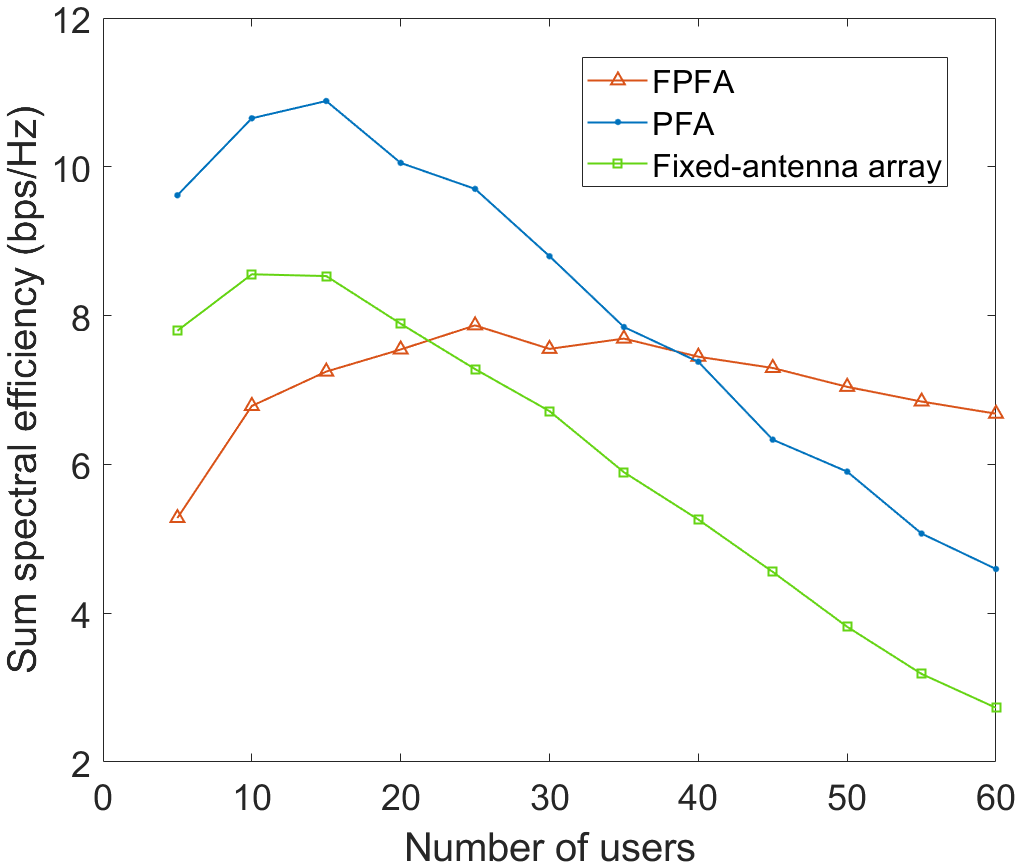}
\caption{Sum SE versus the transmit power: BS with 256 antennas.}
\label{Fig. 256 antennas}
\end{minipage}
\end{figure*}

In this section, we evaluate the performance of the proposed correlation-based frequency allocation strategy, the port selection and the hybrid precoding algorithm in FPFA, PFA, in comparison with the traditional fixed antenna structure. 

\textit{Simulation Setup:} Our simulation considers $K = 4$ frequency bands centered at $270,300,330,360$ GHz with $27,30,33,36$ GHz bandwidth, respectively. Therefore, the BS is divided into $4$ subarrays, each supports one central frequency to serve a group of randomly distributed users. The users are equipped with one PFA or one fixed antenna to support the reception of one data stream. The PFA is able to adjust position among $500$ ports along a fixed line with the length of $15\lambda$. The coverage of a BS ranges up to $25$~m with a beam sector of $120^\circ$. The heights of the BS and UEs are $h_t = 20$~m, and $h_r = 1.5$~m. We consider a multi-path channel with one LoS path and the ground-reflection path, i.e., $N_p = 2$.
The number of transmit RF chains at the FPFA equals $\frac{1}{K}$ of the number of users, i.e., $L_t = \frac{U}{4}$. Since PFA and the fixed antenna system require the number of RF chains no less than the number of users, we set $L_t = U$ in both architectures.

We first assess the sum SE versus the transmit power for a $60$-user scenario in Fig.~\ref{Fig. power}. We observe that the two PFA-enabled systems significantly outperform the fixed-antenna system due to the additional spatial diversity, achieving a sum SE that is over twice that of the fixed-antenna array when transmit power is $30$~dBm. Moreover, the proposed FPFA scheme further improves the sum SE performance with the frequency domain diversity. 

In Fig.~\ref{Fig. EE}, we compare the energy efficiency for the three architectures, which is defined as the ratio between the power consumption and the sum SE~\cite{cluster-based}. The power consumption of each device at the BS is listed in Table \ref{tab: power}. We observe that the energy efficiency decreases with larger number of users. While FPFA maintains notably higher energy efficiency compared to PFA and fixed-antenna arrays, attributed to its capability to reduce the number of RF chains by a factor of $K$.

In Fig.~\ref{Fig. 128 antennas}, we analyze the sum SE versus the number of users, where the BS is equipped with $128$ fixed-position antennas, and the transmit power is set as $20$~dBm. We observe that when the users are few, the PFA architecture has the highest sum SE. This is explained that the user interference can be well eliminated by the spatial multiplexing when they are sparsely located, so that dividing the antennas into subarrays in FPFA is unnecessary. Therefore, in such cases it is preferable to utilize the whole antenna array at the same frequency to provide a high array gain. 
However, as the number of users increases, the user interference becomes severer so that merely exploiting the spatial degree of freedom can no longer fully separate the users. 
As a result, the performance of both PFA and the fixed-antenna architectures experience significant degradation, yet PFA consistently outperforms fixed-antenna array. 
While as the performance of the proposed FPFA architecture remains steady thanks to the frequency domain diversity, it starts to exhibit its superiority over the other two counterparts. 

Fig.~\ref{Fig. 256 antennas} presents the sum SE when the BS is equipped with 256 antennas. Compared with 128 antennas, we observe that the intersection point of FPFA and the PFA shifts to right, i.e., with more users in the network. This illustrates the trade-off between harvesting the array gain and conducting frequency-domain interference cancellation. Specifically, with an increasing number of antennas, PFA and fixed-antenna system are able to support more users due to higher array gain and narrower beams, such that the superiority of FPFA only depicts when the number of users is even larger.

\begin{table}
\centering
\caption{Power consumption at around 300 GHz and hardware components for FPFA, PFA and fixed-antenna array.} \label{tab: power}
\begin{tabular}{|c|c|c|c|c|}
\hline
\multirow{2}*{Devices}& \multirow{2}*{\makecell[c]{Power \\ (mW)}}& \multicolumn{3}{c|}{Number of devices}\\
\cline{3-5}
 & & FPFA & PFA & Fixed-antenna\\
\hline
Phase shifter & 42 \cite{dynamic} & $UN_t/K$ & $UN_t$ & $UN_t$\\
\hline
RF chain & 120 \cite{cluster-based}& U/K & U & U\\
\hline
Baseband & 200 \cite{dynamic} & K & 1 & 1\\
\hline
\end{tabular}
\end{table}

\section{Conclusion}
\label{section_conclusion}
In this paper, the FPFA hybrid beamforming architecture is proposed to enhance the ability of the system to support ultra-dense connectivity. We first establish the channel model and the system model of the FPFA architecture, based on which we show the impact of the port selection and formulate the joint optimization problem to maximize the sum SE.Then, we decompose the problem into three subproblems of frequency allocation, port selection and the hybrid precoding design. For frequency allocation, we divide the users into several frequency groups based on the criterion of minimizing the sum channel correlation coefficient. Next, we derive the minimum-projection-based port selection algorithm to further reduce the inter-user interference within each frequency group. Then, the hybrid precoding scheme is investigated where the digital precoder is obtained through the BD algorithm, and the analog precoder is obtained through SVD of the channel matrix. 

Simulation results show that through the utilization of PFAs, the system performance largely improves due to additional spatial diversity. While the proposed FPFA scheme that can utilize both spatial and frequency diversity exhibits a relatively steady performance with increasing number of users and outperforms PFA and fixed-antenna systems for ultra-dense user deployment. Additionally, with less antennas deployed at the BS, the superiority of the FPFA tends to be more significant.

\bibliographystyle{IEEEtran}
\bibliography{reference}
\end{document}